\documentclass[aps,prl,twocolumn,preprintnumbers,amsmath,amssymb,superscriptaddress]{revtex4-2}
\usepackage{graphicx}
\usepackage[caption=false]{subfig}
\usepackage{epsfig}
\usepackage{dcolumn}
\usepackage{bm}
\usepackage{color}
\usepackage{float}
\definecolor{LinkColor}{rgb}{0.75,0.0,0.2}
\usepackage[colorlinks]{hyperref}
\hypersetup{
    colorlinks=true,
    linkcolor=LinkColor,
    citecolor=LinkColor,
    urlcolor=LinkColor
} 

\begin{document}

\title{Field-induced phase transitions in the Kitaev-Heisenberg model: \\A sign-problem-free quantum Monte Carlo study and possible application to $\alpha$-${\rm RuCl_3}$}

\author{Xuan Zou}
\affiliation{Institute for Advanced Study, Tsinghua University, Beijing 100084, China} 
\affiliation{Department of Physics and Anthony J. Leggett Institute for Condensed Matter Theory, University of Illinois Urbana-Champaign, 1110 West Green Street, Urbana, Illinois 61801, USA}
\author{Shuo Liu}
\affiliation{Institute for Advanced Study, Tsinghua University, Beijing 100084, China} 
\author{Wenan Guo}
\email{waguo@bnu.edu.cn}
\affiliation{School of Physics and Astronomy, Beijing Normal University, Beijing 100875, China}
\affiliation{Key Laboratory of Multiscale Spin Physics (Ministry of Education), Beijing Normal University, Beijing 100875, China}
\author{Hong Yao}
\email{yaohong@tsinghua.edu.cn}
\affiliation{Institute for Advanced Study, Tsinghua University, Beijing 100084, China} 

\date{\today}

\begin{abstract}
The frustrated magnet $\alpha$-${\rm RuCl_3}$ is one of the prime candidates for realizing a Kitaev quantum spin liquid (QSL). However, the existence of a field-induced intermediate QSL phase in this material remains under debate. 
Here, we employ sign-free numerically exact quantum Monte Carlo simulations to investigate the Kitaev-Heisenberg (KH) model on the honeycomb lattice with $K=-2J$ under an applied magnetic field along the $z$-direction. 
Our findings reveal that the system undergoes a direct quantum phase transition from a zigzag magnetically ordered phase to a spin-polarized phase at zero temperature, which belongs to the 3D XY universality class. At finite temperatures, a Berezinskii-Kosterlitz-Thouless transition line separates the spin-polarized phase from a quasi-long-range ordered state, eventually terminating at the quantum critical point.
Our results convincingly show that there is no intermediate QSL phase in the KH model with a $z$-direction magnetic field, which we believe will shed important light on understanding experimental observations in $\alpha$-${\rm RuCl_3}$.
\end{abstract}

\maketitle

{\bf Introduction:}
Quantum spin liquid (QSL) is an exotic quantum phase of matter in which spins do not possess any long-range order even at zero temperature due to strong quantum fluctuations \cite{Anderson1973, Lee2008, Balents2010, Savary2017, Zhou2017, wen2017, Knolle2019, Broholm2020, Norman2016, kivelson2023}.
QSLs cannot be described by the Landau paradigm; instead, they exhibit intriguing features such as fractionalized excitations and topological orders \cite{Lee2008, Balents2010, Savary2017, Zhou2017, wen2017, Knolle2019, Broholm2020, Norman2016, kivelson2023, wen2007, Fradkin2013, Sachdev2023}.
Moreover, QSLs could have promising applications, such as providing a possible mechanism for high-temperature superconductivity \cite{Anderson1987, Kivelson1987, Anderson2004, Lee2006}
and a basis for fault-tolerant quantum computation \cite{Nayak2008, kitaev2003}. 
Theoretically, it was rigorously shown that a QSL ground state can be achieved in various microscopic models, including quantum dimer models and related ones \cite{ Rokhsar1988, Moessner2001, Moessner2001b, Senthil2002, Balents2002} and toric-code models \cite{kitaev2003, Wen2003}.  
In particular, Kitaev introduced an exactly solvable quantum spin model on the honeycomb lattice \cite{Kitaev_2006}, whose ground state is a QSL featuring fractionalized excitations, igniting enormous research interest; see, i.e. Ref. \cite{YaoKivelson2007, Baskaran2008, Mandal2009,  Yao2009, Hong2011, Knolle2014, Yoshitake2016, Rau2016, Banerjee2016, Takagi2019, Lin2021, Zhou2023, Rousochatzakis2024, Hu2024, Zhang2024, ZhengzhiWu2024, Kee-roadmap2024}.

Interestingly, the Kitaev model with a QSL ground state could be realized in materials with strong spin-orbit coupling \cite{Jackeli_2009}, although identifying QSLs in real materials is in general challenging.
So far, a prime candidate for realizing the honeycomb-lattice Kitaev QSL state is the ruthenium-based material $\alpha$-${\rm RuCl_3}$ \cite{Plumb2014, Majumder2015, Sears2015, Banerjee_2016, yuwqprb2017, Ran2017, Kasahara2018, Balz2019, Balz2021, Yokoi2021, Ran_2022, Bruin2022, Czajka2021, Czajka2023, matsuda2025}, although there are other candidate materials \cite{Singh2010, Liu2011, Singh2012, Choi2012, Ye2012, modic2014, Biffin2014, Biffin2014b, hwan_chun2015, Takayama2015, Williams2016, kitagawa2018, Liu2018, Sano2018, Pei2020, Liu2020, Kim2022, Lee2023}.
This material exhibits strong spin-orbit coupling, with spin interactions closely approximating Kitaev physics. 
However, the ground state of $\alpha$-${\rm RuCl_3}$ was experimentally found to show zigzag magnetic order instead of a QSL phase \cite{Johnson2015, Kim2015, Cao2016}. 
The magnetic properties of $\alpha$-${\rm RuCl_3}$ extend beyond the original Kitaev model, necessitating further theoretical and numerical investigations.
Straightforward extensions include the introduction of an additional Heisenberg interaction, resulting in the Kitaev-Heisenberg (KH) model \cite{Jackeli_2009, Chaloupka2010, Chaloupka2013, Li_2015, Gotfryd2017, Gohlke2017, Georgiou2024} and 
an additional $\Gamma$ term \cite{Rousochatzakis2017, Samarakoon2018, Wachtel2019, Gordon2019, Saha2019, Buessen2021, Yamada2020, Gohlke2020, Liu2021, Shi2021, luo2024}. Both the Heisenberg and $\Gamma$ terms are capable of reproducing experimentally observed magnetic zigzag orders with appropriate parameters.

\begin{figure}[t]
    \centering
        \includegraphics[width=0.85 \columnwidth]{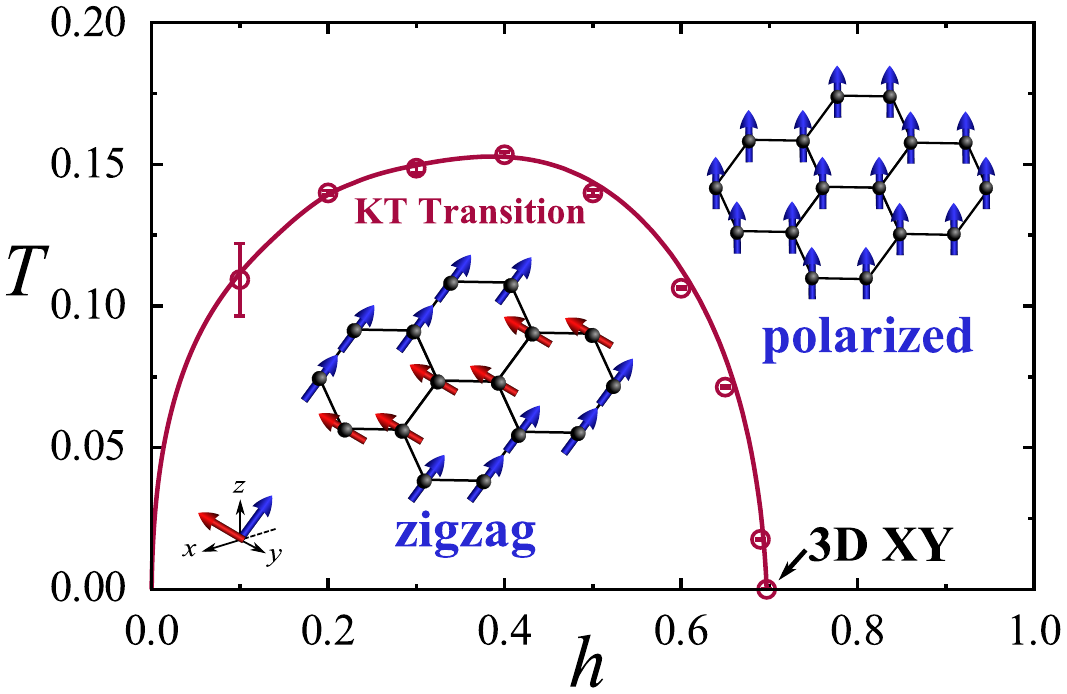}
	\caption{ Phase diagram of the KH model at ($K=2, J=-1$) on the honeycomb lattice under a uniform magnetic field along the $[001]$ direction. $h$ denotes the strength of the magnetic field, and $T$ is the temperature. The thermal phase transition line separating a quasi-long-range ordered state from the spin-polarized state is classified as a BKT transition. 
	At zero temperature, a quantum phase transition in the 3D XY universality class occurs at $h_c \approx 0.697$, where the BKT line terminates. }
\label{afm_phase}
\end{figure}

It is natural to apply a magnetic field to suppress this zigzag magnetic order and potentially induce a non-abelian chiral QSL phase, 
for which a half-quantized value of thermal Hall conductivity would be an 
experimental signature for Majorana excitations in the non-abelian QSL state of the Kitaev model \cite{Kitaev_2006, Nasu2015, Yoshitake2016, Nasu2017, Kumar2023, Fan2023, Kurumaji2023}.  
Although some studies have reported a half-quantized thermal Hall signal \cite{Kasahara2018, Yokoi2021, Bruin2022}, 
the observed quantization appears to depend on sample quality, leaving the occurrence of the field-induced non-abelian QSL state in $\alpha$-${\rm RuCl_3}$ inconclusive \cite{Yamashita2020, Czajka2021, Czajka2023}.
(For a recent review, see Ref \cite{matsuda2025}.)
Numerical studies of these models under an applied magnetic field are expected to provide important insights in interpreting experimental findings \cite{Janssen2016, Chern2017, Jiang2019, Li2021, Kumar2023, Yu2023, bhullar2025fieldinducedorderedphasesanisotropic}. 
However, studies in the density matrix renormalization group (DMRG) suggested the presence of an intermediate QSL phase \cite{Jiang2019, Li2021} only when the magnetic field is oriented out of the plane (in the [111] direction), which appears to contradict experimental results.

In part because controversial conclusions are drawn from experiments and from DMRG calculations on models with magnetic fields on a quasi-1D lattice, it is desired and crucial to perform an unbiased
large-scale quantum Monte Carlo (QMC) study of the models with a magnetic field on a 2D lattice with $L\times L$ geometry. 
However, QMC usually suffers from severe sign problems due to the frustration inherent in the Kitaev interactions.
Fortunately, through a unitary transformation, the KH model with a magnetic field $h$ along the $z$-direction is sign-problem-free at a specific point where the interaction strength coefficients satisfy $K=-2J$ [see Eq. (\ref{equ_H_KH_h1}) below]. In this case, the ground state of this model without a magnetic field is found to be a zigzag state for $K>0$ and a stripy state for $K<0$ \cite{Chaloupka2013}, the former consistent with experimental observations in $\alpha$-${\rm RuCl_3}$. 
For small values of $h$, the in-plane zigzag order persists, which is also consistent with experimental observations; 
while in the strong magnetic field limit, the ground state becomes a state with all spins aligned in the $z$-direction, forming a spin-polarized state. 
A natural question arises whether an intermediate QSL state exists between the zigzag and spin-polarized states, which serves as the primary motivation for our large-scale unbiased QMC study. 

We employ the sign-free stochastic series expansion (SSE) QMC method with directed loop updates\cite{sandvik1991, Sandvik_1992, sandvik2002} 
to investigate the phase diagram of the model with $K=-2J$ under a magnetic field applied along the $z$-direction.  
At zero temperature, we demonstrate that a direct quantum phase transition occurs from the zigzag state ($K>0$) to the polarized state, which belongs to the 3D XY universality class. 
The transition point marks the terminating point of a Berezinskii-Kosterlitz-Thouless (BKT) transition line, which separates the spin-polarized phase from a quasi-long-range ordered state at finite temperature. The phase diagram is illustrated in Fig. \ref{afm_phase}.

\begin{figure}[tbp]
  \centering
  \includegraphics[width=0.95 \columnwidth]{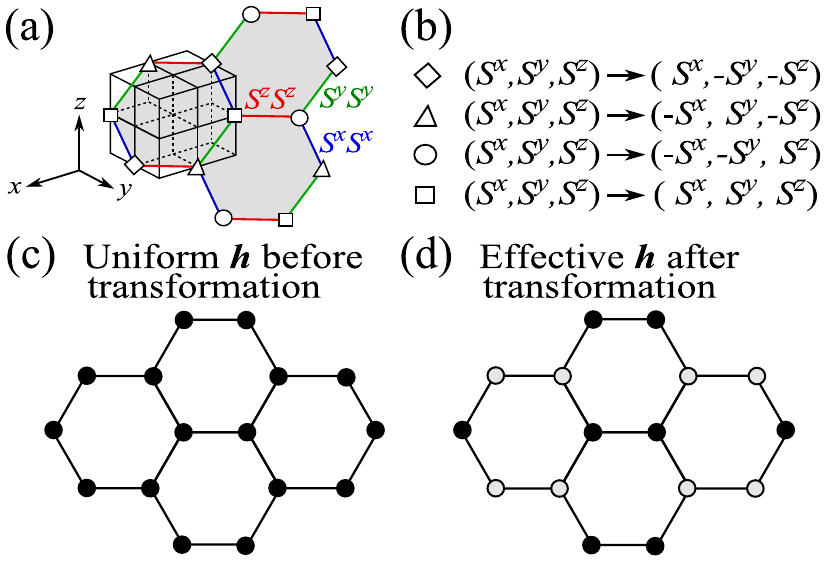}
  \caption{ Transformation of the KH model under a magnetic field.
(a) Visualization of the honeycomb lattice structure in $\alpha$-${\rm RuCl_3}$. The Kitaev term in the Hamiltonian is illustrated by three distinct bond types, each represented by a different color. Lattice sites are marked with four unique shapes, corresponding to the four spin transformation types.
(b) The spin transformation rules illustrating how the spin components $(S^x, S^y, S^z)$ are transformed for each lattice site type.
(c) In the original model, the magnetic field $h$ is uniformly applied along the [001] direction.
(d) In the transformed model, the effective magnetic field $h$ exhibits a stripe-like pattern, with the black dot representing the [001] direction and the light gray dot indicating the [00$\bar{1}$] direction.
} 
  \label{trans}
\end{figure}

{\bf Model:}
The Hamiltonian of the KH model with a uniform magnetic field along the $z$-direction reads
\begin{eqnarray}
\label{equ_H_KH_h1}
H = K\sum_{\langle ij \rangle} S_i^\gamma  S_j^\gamma+J\sum_{\langle ij \rangle}\mathbf{S}_i \cdot \mathbf{S}_j-h\sum_i S_i^z,
\end{eqnarray}
where $S_i^\gamma$ is the spin-1/2 operator on site $i$, 
$K$ is the strength of the well-known Kitaev term [$\gamma=x,y,z$ representing the three types of nearest-neighbor bonds in the honeycomb lattice, as illustrated in Fig. \ref{trans}(a)], the $J$ term corresponds to Heisenberg interactions of two spins on neighboring sites, and the last term is the Zeeman coupling induced by the magnetic field along the $z$-direction.
At $J=h=0$, this model reduces to the Kitaev model, with the ground state being the gapless quantum spin liquid. 
Experimental results for the Kitaev material
$\alpha$-${\rm RuCl_3}$ show that its ground state exhibits zigzag order \cite{Johnson2015, Kim2015, Cao2016}, indicating the existence of other interactions such as the Heisenberg interactions that could help induce such magnetic ordering.   
Turning on the magnetic field term suppresses the magnetic order and may induce a quantum spin liquid phase.

For the KH model without a magnetic field ($h=0$), the ground state exhibits a zigzag order within the finite range $-0.04< J/K < -1.15$ \cite{Georgiou2024}.  
Throughout this study, we fix the interaction strength coefficients to $K=2$ and $J=-1$ for the following reason. 
For this specific parameter choice, the KH model in the honeycomb lattice can be mapped onto the Heisenberg model through a transformation \cite{Chaloupka2010}, as illustrated in Fig. \ref{trans}(b). 
Before the transformation, the KH model is subject to a uniform magnetic field $h$, as shown in Fig. \ref{trans}(c). 
After applying the transformation, the Hamiltonian takes the form: $H = K/2 \sum_{\langle ij \rangle}\mathbf{S}_i \cdot \mathbf{S}_j-\sum_i h_i S_i^z$.
\begin{figure}[tp]
    \centering
        \includegraphics[width=1.0 \columnwidth]{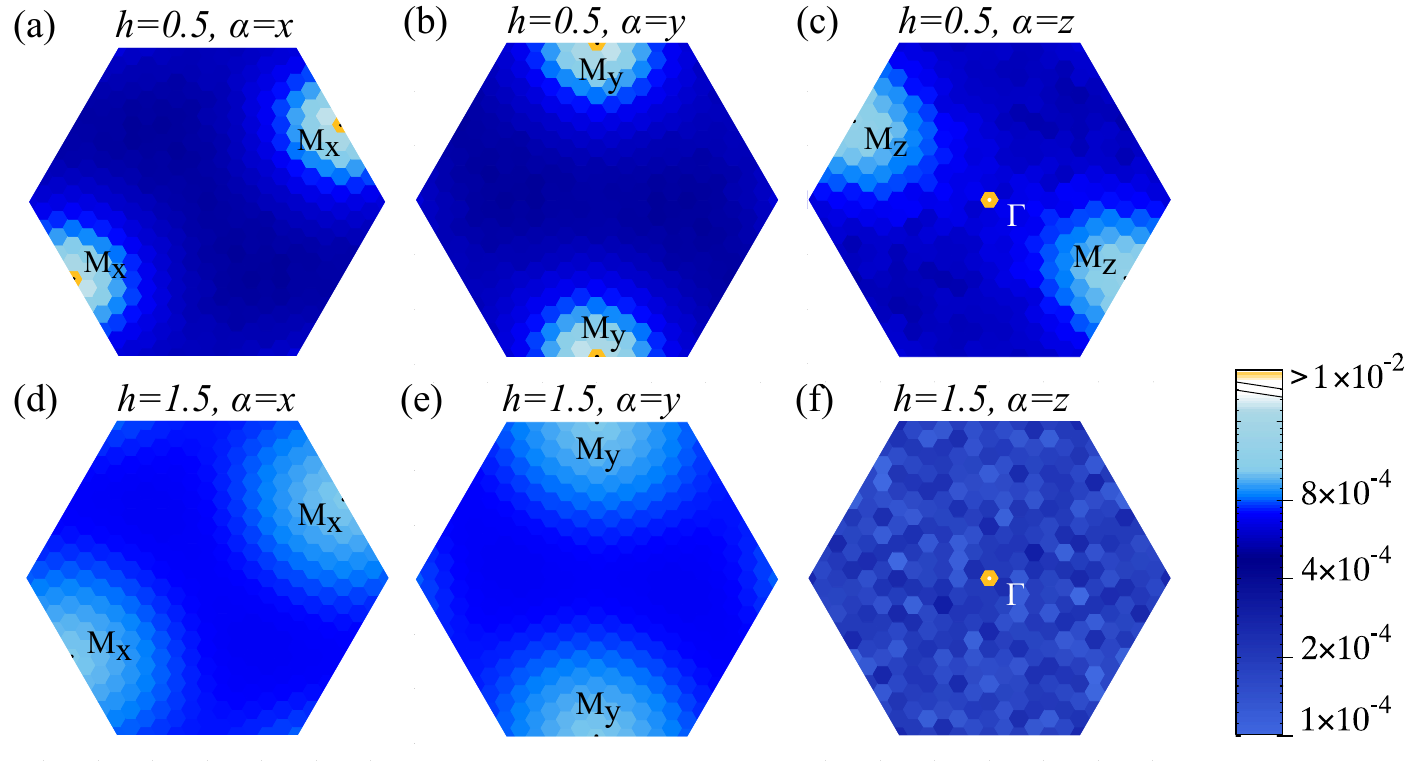}
        \caption{
Structure factor $M_{2}^{x,y,z}(\vec{Q})$ in the first Brillouin zone: comparing zigzag state ($h=0.5$) and polarized state ($h=1.5$) for system size $L=18$ at inverse 
temperature $\beta=L$. 
(a-c) The zigzag state at $h=0.5$, with $M_{2}^{x,y,z}(\vec{Q})$ showing peak values at $M_{x}$, $M_y$, and $\Gamma$ points, respectively. 
The peak of $M_{2}^{z}$ at $\Gamma$ is induced by the $z$-direction field, while the zigzag order is an in-plane ordered state.
(d-f) The spin-polarized state at $h=1.5$, where $M_{2}^{z}(\vec{Q})$ exhibits a peak at the $\Gamma$ point, and the in-plane zigzag order vanishes. }
\label{afm_stru}
\end{figure}
Note that the direction of the magnetic field in transformed $H$ is site-dependent, alternating between $+h$ and $-h$, resulting in a stripe-like pattern, as shown in Fig. \ref{trans}(d).
In other words, the original KH model subject to a uniform magnetic field [Fig. \ref{trans}(c)] is transformed into the pure Heisenberg model under a stripe-patterned magnetic field [Fig. \ref{trans}(d)].
This transformation enables us to conduct quantum Monte Carlo simulations without encountering notorious sign problems.

Before delving into the numerical calculations, we examine the ground states in the limits of zero magnetic field and strong magnetic field, respectively.
At $h=0$, the model reduces to an antiferromagnetic (AFM) Heisenberg model on a bipartite lattice. As a result, the ground state displays SU(2) symmetry breaking, leading to the formation of a N\'eel state. In this case, no finite-temperature phase transition is expected. Applying the inverse transformation yields the ground state of the original KH model, which naturally adopts the zigzag ordered state, as depicted in Fig. \ref{afm_phase}. 
Applying small values of $h$ breaks the spin SU(2) rotations and results in the in-plane zigzag order, which is consistent with experimental observations of the in-plane zigzag order in $\alpha$-${\rm RuCl_3}$.
On the other hand, in the strong magnetic field limit, the ground state of the transformed Heisenberg model becomes a state with all spins aligned in the direction of the transformed field $h_i$, forming a stripe pattern, which corresponds to all spins in the original KH model aligned in the $z$-direction, forming a spin-polarized state. 

One central question to ask is whether an intermediate QSL emerges between the zigzag magnetic order at low field and the spin-polarized state at high field, or whether a direct phase transition occurs between the two states. To address this question, we perform sign-problem-free QMC simulations to systematically study the phase diagram as the strength of the magnetic field is varied. 
We will demonstrate that no intermediate spin liquid state exists between the zigzag and spin-polarized states. Instead, there is a direct
transition that belongs to the 3D XY universality class.

\begin{figure}[tbp]
    \centering
        \includegraphics[width=0.98 \columnwidth]{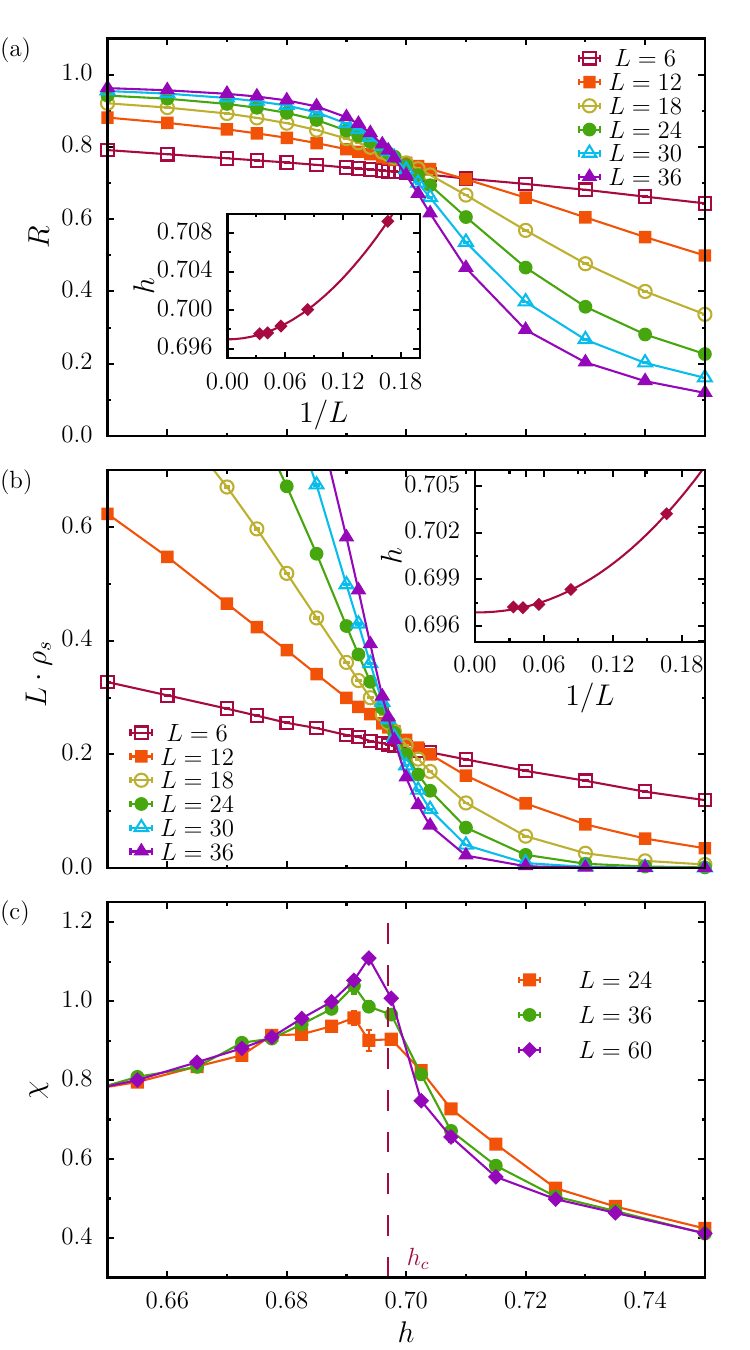}
         \caption{
Phase transition at zero temperature in the KH model with a uniform magnetic field. (a) The RG invariant ratio of zigzag order shows crossing points, indicating the transition at critical magnetic field $h_c=0.6970(1)$. Inset: an FSS analysis of $h_c$. (b) The same transition point, $h_c=0.6969(1)$, is obtained from $L\rho_s$, consistent with the RG invariant ratio. Inset: an FSS analysis of $h_c$. (c) Magnetic susceptibility $\chi={\rm d} M^z/{\rm d}h$ exhibits a single peak around the quantum phase transition point, suggesting the absence of an intermediate spin liquid phase. The dashed line marks $h_c=0.697$.
         }
\label{xbinder}
\end{figure}
{\bf Numerical results:}
Using QMC simulations, we identify the phases by evaluating the structure factors, which are defined as the Fourier transform of the correlation function:
$M_{2}^\alpha(\vec{Q})=(1 / N^{2}) \sum_{i, j }\langle S_{i}^{\alpha} S_{j}^{\alpha}\rangle e^{i \vec{Q} \cdot\left(\vec{r}_{i}-\vec{r}_{j}\right)}$,
where $\alpha=x, y, z$, and $N$ is the number of unit cells, and $\vec{Q}$ the crystal momentum.

We begin by studying the ground state of the model. This is achieved by setting the inverse temperature $\beta=L$, where $L$ is the linear size of the system, and performing finite-size scaling (FSS) to determine properties as $L\to \infty$. 
Fig. \ref{afm_stru} shows the distribution of three components of the structure factors $M_{2}^{x,y,z}(\vec{Q})$ in the first Brillouin zone
for a system of size $L=18$ at two typical magnetic field values $h=0.5$ and $h=1.5$. 
For $h=0.5$, $M_{2}^{x,y,z}(\vec{Q})$ peaks at $\vec Q=\vec{Q}^{*}$ with $\vec Q^*=M_{x}$, $M_y$, or $\Gamma$ points,  illustrating the system bearing an in-plane zigzag (or canted zigzag) order for sufficiently low temperatures, as shown in Fig. \ref{afm_stru}(a-c). This in-plane zigzag state is consistent with results from a previous large-$S$ study \cite{Janssen2016}. 
For $h=1.5$,  as shown in Fig. \ref{afm_stru}(d-f), only $M_{2}^{z}(\vec{Q})$ exhibits a peak at the $\Gamma$ point, indicating the system is in a spin-polarized state. 

To study whether there is an intermediate spin liquid state between these two states, we calculate the renormalization-group invariant ratio, which is defined as
$R(h, L)=1-M_{2}^\alpha (\vec{Q}^{*}-\delta \vec{Q})/M_{2}^\alpha (\vec{Q}^{*})$ with $\alpha=x$.
Here, $\vec{Q}^*$ denotes the ordering wave vector for the corresponding spin component $\alpha$, 
while $\delta \vec{Q}$ represents the minimum crystalline momentum interval. 
In the ordered phase, the ratio $R$ approaches $1$ as the system size $L$ increases to infinity, while it tends to zero in the disordered phase. Suppose a critical point $h_c$ separates the two phases; since $R(h, L)$ is dimensionless at $h_c$, $R(h, L)$ for different system sizes across near $h_c$, roughly indicating the transition point. This is the case shown in Fig. \ref{xbinder}(a). 

The spin stiffness $\rho_s$ is another observable that
provides information about phase transitions at both zero and finite temperatures. 
It is calculated through the average fluctuations of the winding number of spin transport, $\langle W_\alpha ^2 \rangle/\beta$ along the $\alpha=x$ and $y$ directions, 
where $\beta$ is the inverse temperature \cite{Pollock1987, sandvik2010}. For a 2D quantum critical point with dynamical exponent $z=1$, $L \rho_s$ is dimensionless at the critical point \cite{Fisher_boson}. 
We observe that $L \rho_s(h, L)$ for different system sizes cross near $h_c$, indicating a transition, as shown in Fig. \ref{xbinder}(b).

To precisely locate the transition point and determine the transition properties, we use the finite-size scaling of a quantity $A$ with the scaling dimension $X_A$~\cite{Nightingale, Fisher_Baber}, 
\begin{equation}
    A(\delta, L) = L^{-X_A} f(\delta L^{1/\nu}, L^{-\omega}),
    \label{fss}
\end{equation}
where $\delta$ is the distance to a critical point, and $\nu$ is the correlation length exponent. 
At the critical point, $L^{X_A} A$ becomes dimensionless.
The function $f(x,y)$ is a scaling function with corrections to scaling included, where $\omega > 0$ is the effective exponent for these corrections.

We adopt the standard $(L, 2L)$ crossing analysis to estimate the critical point; for details, see, e.g., Ref. \cite{Shao_2016}.
For the case that $A$ denotes the RG invariant ratio $R$, we have $X_A=0$. 
The crossings $h_c(L)$ of $R(h, L)$ and $R(h, 2L)$ are expected to converge to the critical point $h_c$ in a power law $h_c(L)=h_c+c L^{-x}$, 
where $c$ is a non-universal constant, and $x=1/\nu+\omega$. 
By fitting this power law to the estimated $h_c(L)$, we obtain $h_c=0.6970(1)$.
For the case where $A$ denoting $\rho_s$, $X_A=1$, and $L\rho_s$ is dimensionless.
We apply similar analyses for $L\rho_s(h, L)$ and find $h_c=0.6969(1)$. 
Remarkably, the analyses of $R$ and $L \rho_s$ yield consistent estimates of the critical point $h_c$. 
The insets in Fig. \ref{xbinder}(a,b) show the results of this analysis.

\begin{figure}[tbp]
    \centering
        \includegraphics[width=1.0 \columnwidth]{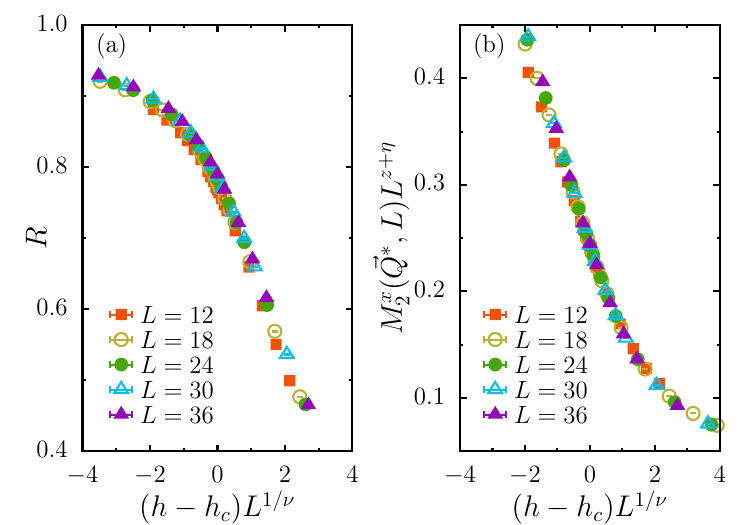}
         \caption{ Data collapse  of (a) $R(h, L)$ and (b) $M_2^x(\vec{Q}^{*},L)$ using $h_c=0.6970$ and 3D XY critical exponents $\eta=0.03810$ and $\nu=0.67169$ \cite{Hasenbusch3dxy}.
        }
\label{afm_collapse}
\end{figure}

We further investigate the universal properties of the critical point using data collapse. 
Fig. \ref{afm_collapse} demonstrates a compelling data collapse of $R(h, L)$ and the structure factor $M^x_2(\vec{Q}^{*}, L)$ near $h_c=0.6970$, utilizing the 3D XY critical exponent $\eta=0.03810(8)$ and $\nu=0.67169(7)$ \cite{Hasenbusch3dxy, Dengyj3dxy}. 
For $M_2^x(\vec{Q}^{*})$, which corresponds to $A$ in Eq. (\ref{fss}), we use $X_A=z+\eta$ with $z=1$. 
This data collapse provides strong evidence that the single transition from the in-plane zigzag magnetic ordered state to the spin-polarized state 
belongs to the 3D XY universality class.

To further demonstrate that there is only one phase transition between the zigzag phase and the polarized phase, we study the magnetic susceptibility, defined as the derivative of uniform magnetization $M^z$ with respect to the magnetic field $\chi={\rm d} M^z/{\rm d}h$. If an intermediate QSL phase exists, $ \chi(h, L)$ would be expected to exhibit a suppression signature within the phase \cite{Jiang2019, Li2021}. 
However, as shown in Fig. \ref{xbinder}(c), $\chi(h, L)$ for different system sizes exhibits a single peak near $h_c=0.697$, indicating a direct transition from the zigzag state to the polarized state.

We now turn to determining the finite-temperature phase diagram. As discussed previously, for $h=0$, the transformed Hamiltonian exhibits SU(2) symmetry, and no finite-temperature phase transition is expected. 
With a small magnetic field applied, the ground state is an in-plane zigzag state, which spontaneously breaks the remaining U(1) symmetry. At low temperatures, although there is no long-range order due to the Mermin-Wegner theorem, the spin stiffness remains finite.
As the temperature increases to the critical point $T_{\rm BKT}$, a topological BKT phase transition occurs \cite{Berezinskii1970, Kosterlitz1973, Kosterlitz1974}. At this transition, unbound vortex-antivortex pairs proliferate due to thermal fluctuations, causing the spin stiffness $\rho_s$ to drop to zero. According to the Nelson-Kosterlitz relation \cite{nelson_universal_1977}, the spin stiffness at the transition satisfies $\rho_{\mathrm{s}}\left(T_{\mathrm{BKT}}\right) = 2 T_{\mathrm{BKT}} / \pi$.

For a finite-size system, we define the finite-size critical temperature $T^*(L)$ as the temperature satisfying $\rho_s(T^*, L)=2T^*/\pi$.
This temperature converges to $T_{\rm BKT}$ at $L\to \infty$ in the form of
$T^{*}(L)=T_{\mathrm{BKT}}+a/\ln ^{2}(b L)$, where $a$ and $b$ are non-universal constants \cite{hsieh_finite-size_2013}.
By calculating $\rho_s(T,L)$ at different values of $h$ and 
fitting the formula to the estimated $T^*(L)$ for various $h<h_c$, we obtain
$T_{\rm BKT}$ for the corresponding $h$, as illustrated in the phase diagram (see Fig.~\ref{afm_phase}). More details can be found in the supplemental materials~\footnote{see the Supplemental Materials for finite-temperature behavior at $K=2, J=-1$; results
at $K=-2, J=1$; and details of the QMC method applied to this study.}.

{\bf Conclusion and Discussion:}
In conclusion, we have investigated the phase diagram of the honeycomb KH model with interaction strength coefficients satisfying
the specific condition $J/K=-1/2$ under a uniform magnetic field, using large-scale quantum Monte Carlo simulations.
We found that there is a direct phase transition from an in-plane zigzag state to a spin-polarized state for $K>0$, without an intermediate QSL phase. 
Through finite-size scaling analyses of the structure factors and spin stiffness, we determined the location and critical properties of the zero-temperature quantum critical point. 
The quantum phase transition was verified to belong to the 3D XY universality class.
Additionally, we studied the finite temperature transitions leaving a quasi-long-range ordered phase, which connects to the zigzag-ordered state at zero temperature, to the spin-polarized phase. The transitions were shown to be BKT transitions, and the transition temperatures were determined for various $h$. 

Although our study is restricted to the parameter space of $J/K=-1/2$, 
our results clearly support the experimental conclusion that there might be no QSL ground state in such a system with a [001] magnetic field. 
In other words, a magnetic field with a direction different from [001], i.e. an in-plane field, or other interactions such as the Gamma interactions may be necessary to realize an intermediate spin-liquid state, which requires further study in the future. 

~\newline
\textbf{Acknowledgements:}
We thank Wen-Jing Zhu and Hae-Young Kee for the helpful discussions. This work was supported in part by NSFC under Grant Nos. 12347107 (XZ, SL, and HY), 12175015 (WG), and 12334003 (HY), by MOSTC under Grant No. 2021YFA1400100 (HY), and by the Xplorer Prize through the New Cornerstone Science Foundation (HY). 
X. Z. acknowledges the support of the Tsinghua Visiting Doctoral Students Foundation and the hospitality of the AJL-ICMT at the University of Illinois Urbana-Champaign.

\bibliographystyle{apsreve}
\bibliography{main}

%\clearpage

\begin{widetext}
    \section*{Supplemental Material}
    \renewcommand{\theequation}{S\arabic{equation}}
    \setcounter{equation}{0}
    \renewcommand{\thefigure}{S\arabic{figure}}
    \setcounter{figure}{0}

\subsection{A. Additional Results of the KH model at $K=2, J=-1$: finite-temperature behavior }

In the main text, we primarily focused on field-induced phase transitions at zero temperature. In this Supplemental Section, we present additional numerical results on finite-temperature behavior, specifically examining the spin stiffness, which is pivotal for determining the finite-temperature phase diagram presented in Fig. \ref{afm_phase}.

\begin{figure}[htbp]
    \centering
        \includegraphics[trim=10 0 10 0, clip, width=6.7cm,keepaspectratio]{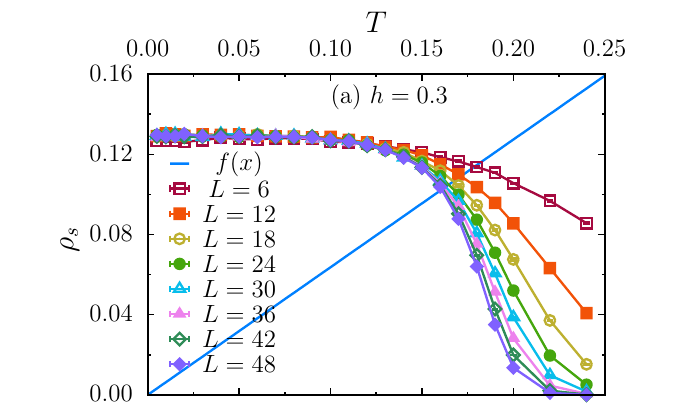}
        \hspace{-0.4cm}
        \vspace{-0.2cm}
        \includegraphics[trim=10 0 10 0, clip, width=6.7cm,keepaspectratio]{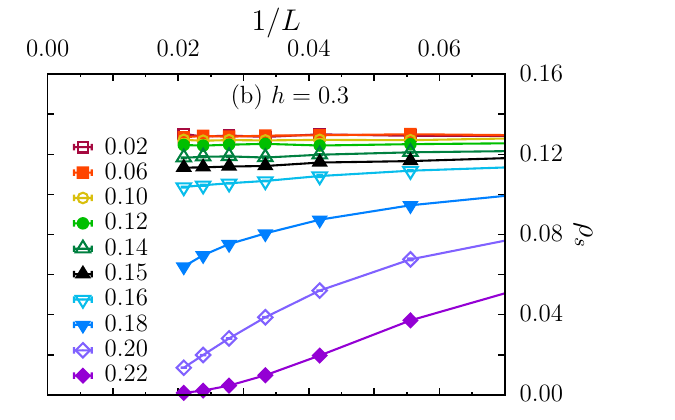}
        \includegraphics[trim=15 0 10 0, clip, width=6.7cm,keepaspectratio]{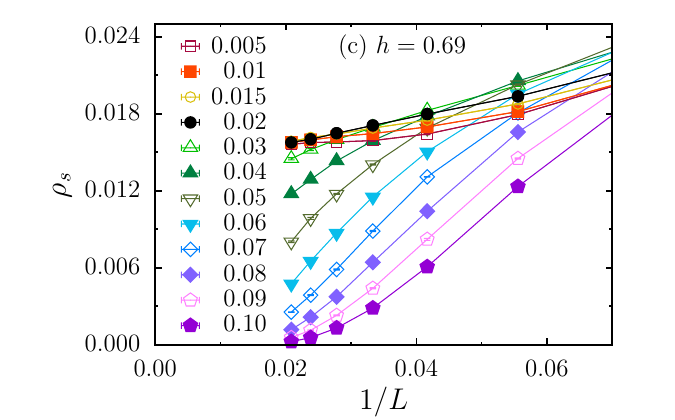}
         \hspace{-0.4cm}
        \includegraphics[trim=8 0 15 0, clip, width=6.7cm,keepaspectratio]{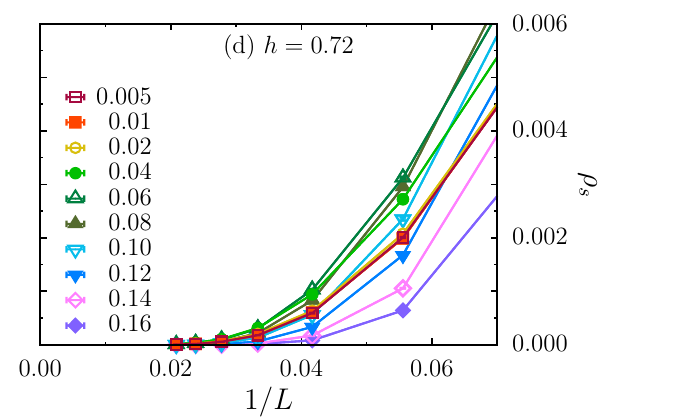}
        \caption{Spin stiffness of the KH model at $K=2, J=-1$ across various temperatures and system sizes. (a) For a fixed magnetic field of $h=0.3$, the spin stiffness exhibits a universal jump at the critical temperature $T_{\rm BKT}$ as the system size increases. The blue line represents $\rho_s(T) = 2T/\pi$. (b-d) Scaling behavior of the spin stiffness across a range of temperatures for different magnetic fields. The corresponding magnetic field $h$ and critical temperature $T_c$ are: (b) $h=0.30$, $T_c = 0.149(2)$, (c) $h=0.69$, $T_c = 0.0176(2)$, and (d) $h=0.72$, $T_c = 0$.}
\label{afm_stf}
\end{figure}

As shown in Fig. \ref{afm_stf}(a), the spin stiffness exhibits a universal jump at the critical temperature $T_{\rm BKT}$, with this behavior becoming more pronounced as the system size increases. Fig. \ref{afm_stf}(b-d) further illustrates the scaling behavior of the spin stiffness across various temperatures as the system size grows. Above the critical temperature $T_c$, the stiffness gradually diminishes to zero, while it retains finite values below $T_c$. These results are consistent with the characteristic features of the BKT phase transition.

\subsection{B. Numerical Results of the KH Model at $K=-2, J=1$}

In the main text, we discussed that the KH model is sign-problem-free at $K=-2J$ and focused primarily on phase transitions at $K=2, J=-1$. In this Supplemental Section, we explore the alternative case of $K=-2, J=1$.

The phase diagram of the KH model on the honeycomb lattice for $K=-2, J=1$ is presented in Fig. \ref{sm_fm_phase}. Our results indicate a direct phase transition from a stripy state to a spin-polarized state at zero temperature, which belongs to the 3D XY universality class. At finite temperatures, the system undergoes a BKT transition.

    \begin{figure}[H]
        \centering
        \includegraphics[trim=0 0 0 0, clip, width=10.0cm,keepaspectratio]{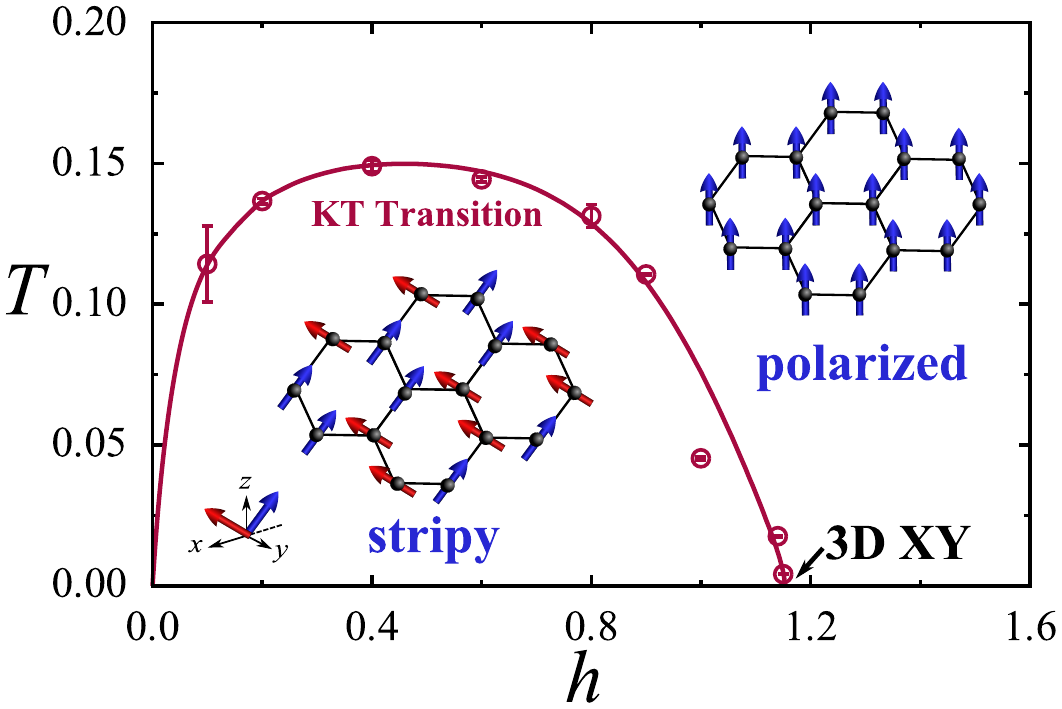}
        \caption{Phase diagram of the KH model at ($K=-2, J=1$) on the honeycomb lattice under a uniform magnetic field along the $[001]$ direction. $h$ denotes the strength of the magnetic field, and $T$ is the temperature. The thermal phase transition from the stripy state to the spin-polarized state is classified as a BKT transition. 
	At zero temperature, a quantum phase transition in the 3D XY universality class occurs at $h_c \approx 1.153$. 
    }
    \label{sm_fm_phase}
    \end{figure}

Fig.~\ref{fm_stru} shows the distribution of the three components of the structure factors, $M_{2}^{x,y,z}(\vec{Q})$, in the first Brillouin zone for a system of size $L=18$ at two representative magnetic field values: $h=0.5$ and $h=1.5$. 

For $h=0.5$, $M_{2}^{x,y,z}(\vec{Q})$ peaks at $\vec{Q}^{*} = M_x, M_y$, and $\Gamma$ points, respectively. Notably, the peak locations are consistent with those reported for $K=2, J=-1$ in the main text. By analyzing the pattern of the state in the transformed Heisenberg model and applying the inverse transformation, it becomes evident whether the ground state is a zigzag state or a stripy state. In this case, the system at $K=-2, J=1$ exhibits an in-plane stripy (or canted stripy) order at sufficiently low temperatures, as shown in Fig.~\ref{fm_stru}(a-c). This in-plane stripy state aligns well with the results of a previous large-$S$ study \cite{Janssen2016}.
For $h=1.5$, as shown in Fig. \ref{fm_stru}(d-f), only $M_{2}^{z}(\vec{Q})$ exhibits a peak at the $\Gamma$ point, signifying a spin-polarized state.

    \begin{figure}[htbp]
        \centering
            \includegraphics[trim=0 0 0 0, clip, width=12cm,keepaspectratio]{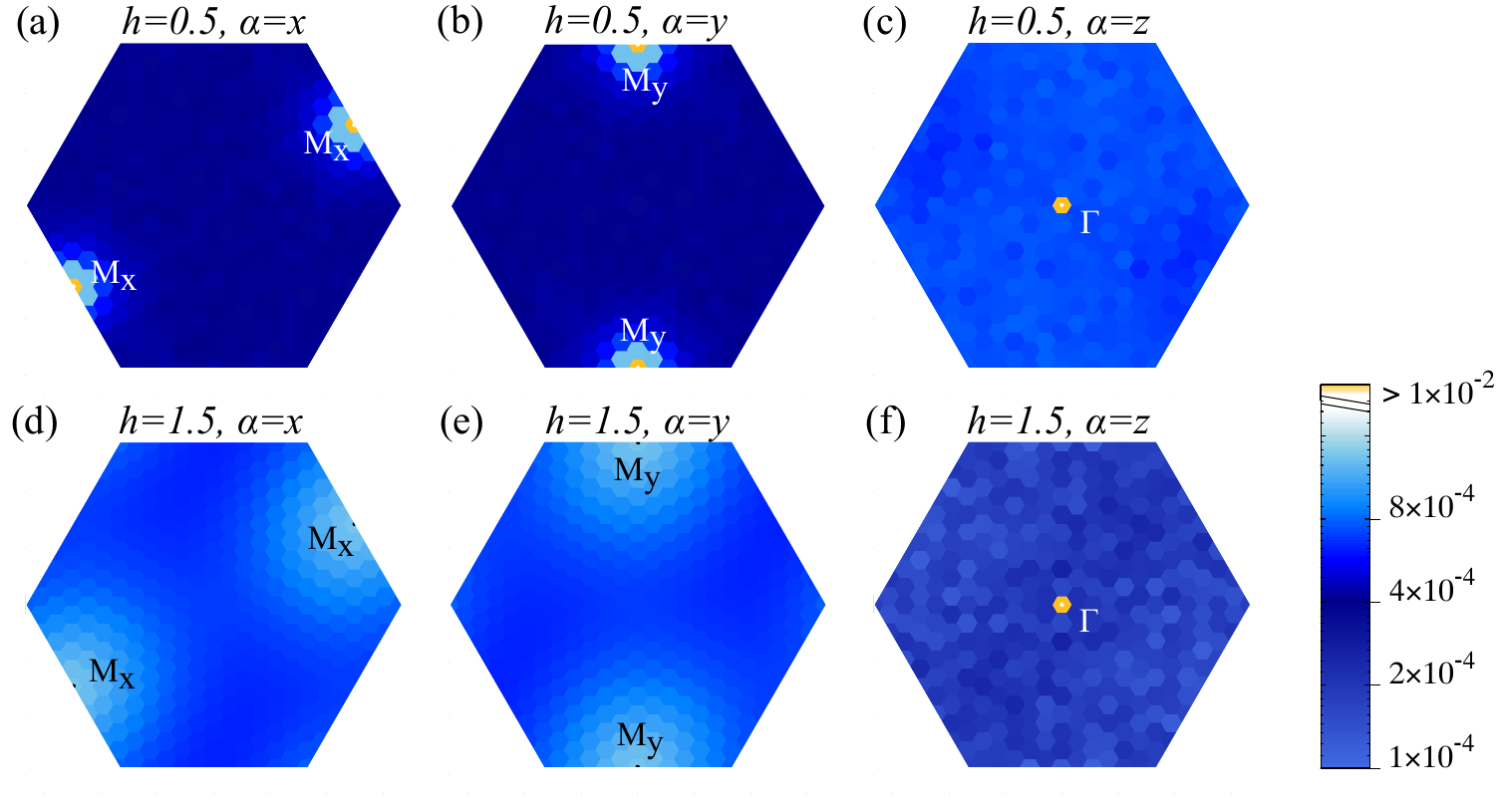}
            \caption{Structure factor $M_{2}^{x,y,z}(\vec{Q})$ in the first Brillouin zone: comparing stripy state ($h=0.5$) and polarized state ($h=1.5$) for system size $L=18$ at inverse temperature $\beta=L$. (a-c) The stripy state at $h=0.5$, with $M_{2}^{x,y,z}(\vec{Q})$ showing peak values at $M_{x}$, $M_y$, and $\Gamma$ points, respectively. The peak of $M_{2}^{z}$ at $\Gamma$ is induced by the $z$-direction field, and the stripy order is an in-plane ordered state. (d-f) The spin-polarized state at $h=1.5$, where $M_{2}^{z}(\vec{Q})$ exhibits a peak at the $\Gamma$ point, and the in-plane stripy order vanishes. 
            }
    \label{fm_stru}
    \end{figure}

To investigate the potential existence of an intermediate spin liquid state between the stripy and polarized states, we calculate the RG invariant ratio and spin stiffness, as shown in Fig.~\ref{sm_fm_xbinder}(a)(b). Finite-size scaling is then employed to precisely locate the transition point and determine the properties of the transition. 
By fitting a power law to the estimated $h_c(L)$, we obtain $h_c = 1.1531(4)$ from the RG invariant ratio and $h_c = 1.1531(1)$ from the spin stiffness, yielding consistent estimates for the critical point $h_c$. 

As in the main text, we analyze the magnetic susceptibility, as shown in Fig.~\ref{sm_fm_xbinder}(c). The susceptibility $\chi(h, L)$ for different system sizes exhibits a single peak near $h_c = 1.1531$, confirming a direct transition from the stripy state to the polarized state.

We further investigate the universal properties of the critical point through data collapse. Fig.~\ref{fm_collapse} presents a compelling data collapse of $R(h, L)$ and the structure factor $M_2(\vec{Q}^{*}, L)$ near $h_c = 1.1531$, using the 3D XY critical exponents $\eta = 0.03810(8)$ and $\nu = 0.67169(7)$~\cite{Hasenbusch3dxy, Dengyj3dxy}. 
This successful data collapse provides robust evidence that the single transition from the in-plane stripy magnetic ordered state to the spin-polarized state belongs to the 3D XY universality class.
    
    \begin{figure}[H]
        \centering
        \includegraphics[trim=0 0 10 0, clip, width=5.6cm,keepaspectratio]{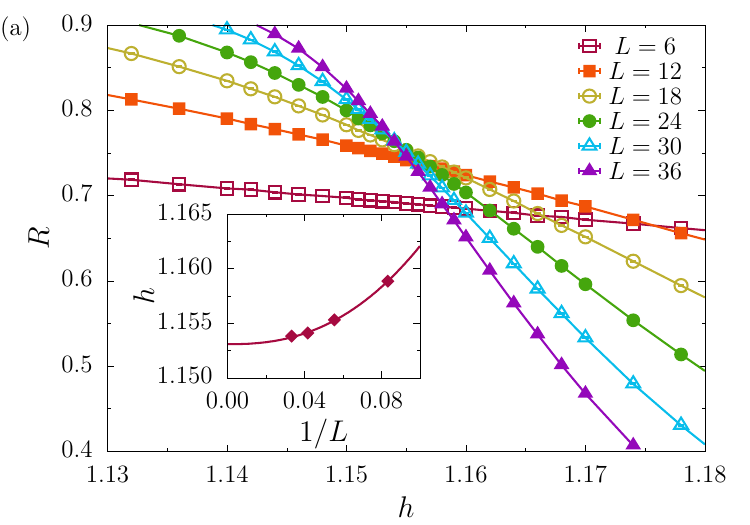}
        \includegraphics[trim=0 0 10 0, clip, width=5.6cm,keepaspectratio]{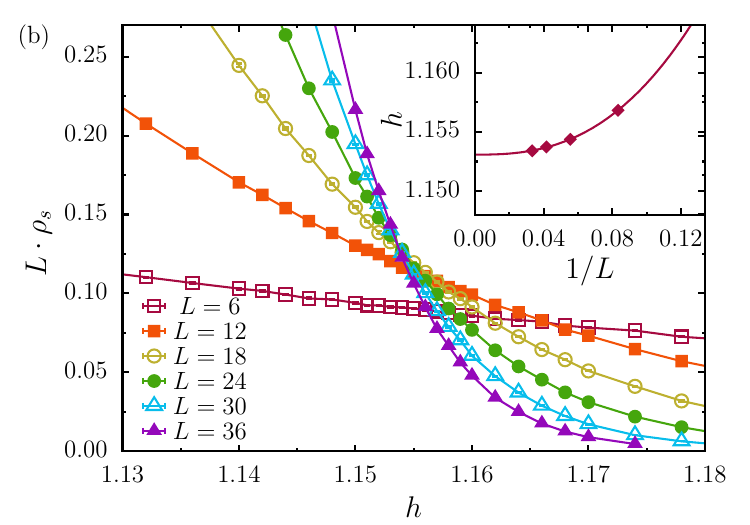}
        \includegraphics[trim=0 0 10 0, clip, width=5.6cm,keepaspectratio]{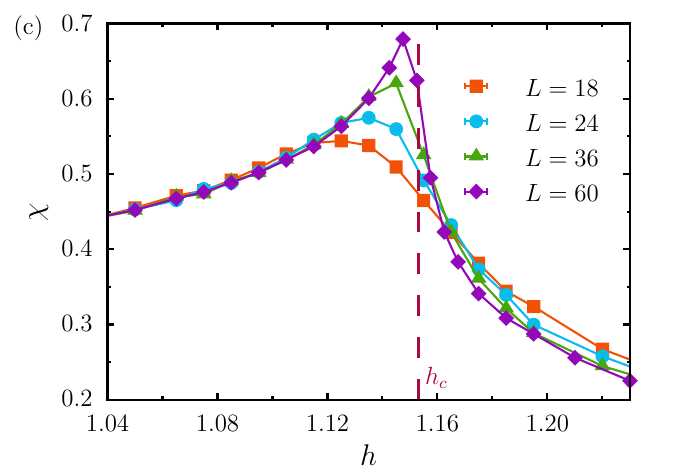}
         \caption{
         Phase transition at zero temperature in the KH model at $K=-2, J=1$ with a uniform magnetic field. (a) The RG invariant ratio of stripy order shows crossing points indicating the transition at critical magnetic field $h_c=1.1531(4)$. Inset: Finite-size scaling of the critical magnetic field. (b) The same transition point obtained from  $L\rho_s$, $h_c=1.1531(1)$, consistent with the RG invariant ratio. Inset: Finite-size scaling result. (c) Magnetic susceptibility $\chi={\rm d} M^z/{\rm d}h$ exhibits a single peak around the quantum phase transition point, suggesting the absence of an intermediate spin liquid phase. The dashed line marks $h_c=1.1531$.
         }
    \label{sm_fm_xbinder}
    \end{figure}

    \begin{figure}[htbp]
        \centering
            \includegraphics[trim=0 0 0 0, clip, width=9cm,keepaspectratio]{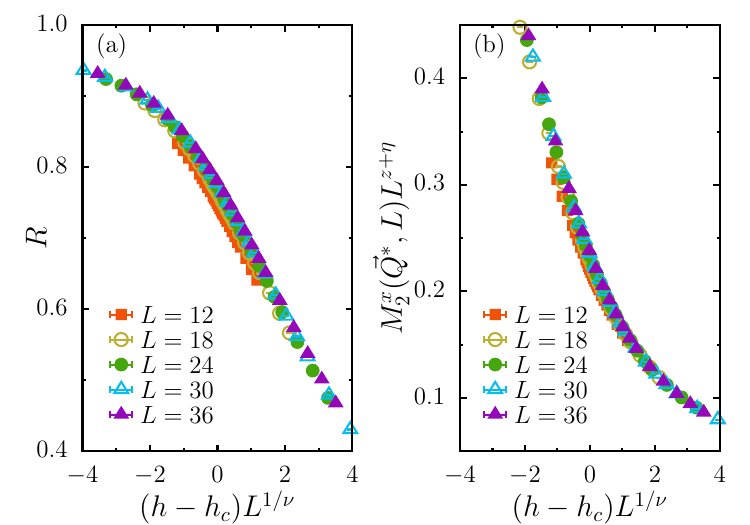}
             \caption{ 
             Data collapse  of (a) $R(h, L)$ and (b) $M_2(\vec{Q}^{*},L)$ using $h_c=1.1531$ and 3D XY critical exponents $\eta=0.03810$ and $\nu=0.67169$ \cite{Hasenbusch3dxy}.
             }
    \label{fm_collapse}
    \end{figure}

We now turn to the finite-temperature transition behavior, focusing on the spin stiffness, which is essential for constructing the phase diagram shown in Fig.~\ref{sm_fm_phase}.
As depicted in Fig. \ref{sm_fm_stf}(a), the spin stiffness exhibits a universal jump at the critical temperature $T_{\rm BKT}$, with this characteristic becoming more distinct as the system size grows. Fig. \ref{sm_fm_stf}(b-d) illustrates its scaling behavior over a range of temperatures, where the stiffness vanishes above the critical temperature $T_c$ but remains finite below it. These results align with the well-established properties of the BKT phase transition.

    \begin{figure}[H]
        \centering
        \includegraphics[trim=0 0 10 00, clip, width=7cm,keepaspectratio]{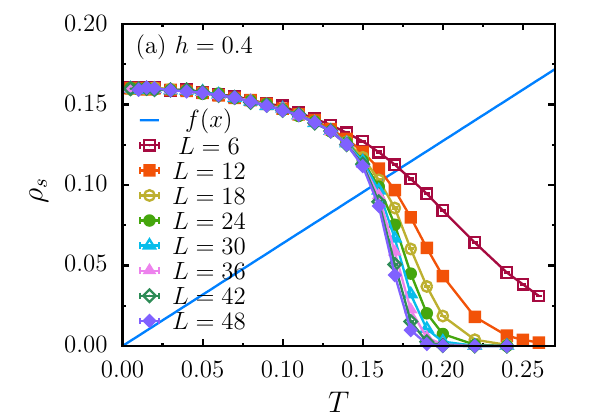}
        \includegraphics[trim=0 0 10 00, clip, width=7cm,keepaspectratio]{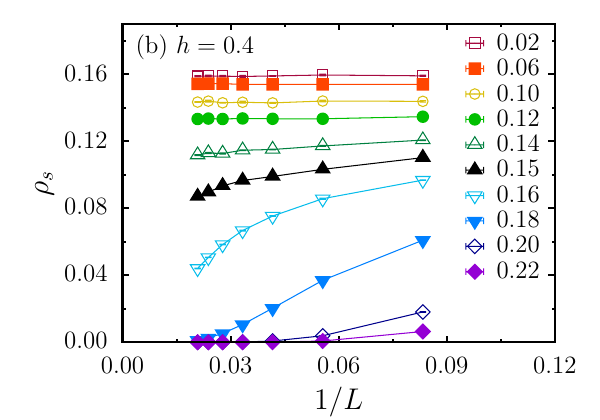}
        
        \includegraphics[trim=10 0 10 00, clip, width=7cm,keepaspectratio]{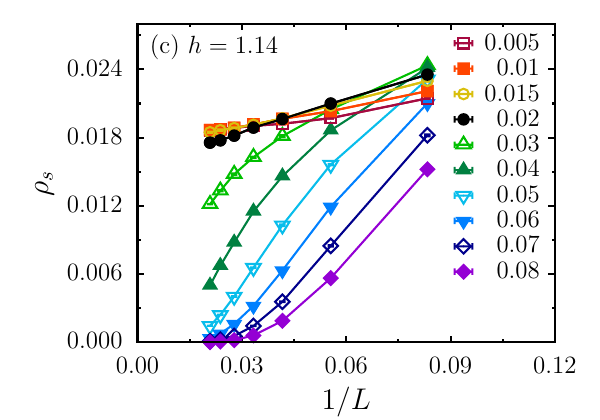}
        \includegraphics[trim=10 0 10 00, clip, width=7cm,keepaspectratio]{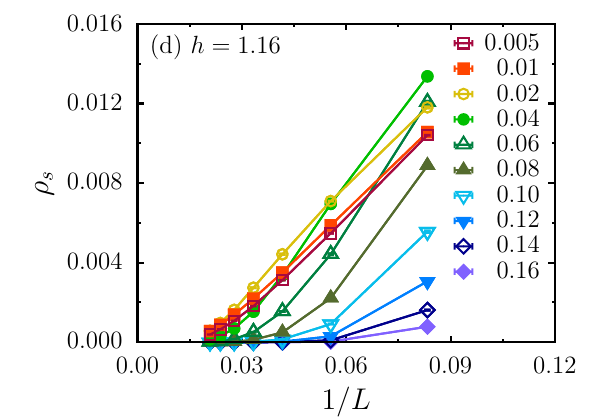}
        \caption{
        Spin stiffness of the KH model at $K=-2, J=1$ across various temperatures and system sizes. (a) For a fixed magnetic field of $h=0.4$, the spin stiffness exhibits a universal jump at the critical temperature $T_{\rm BKT}$ as the system size increases. The blue line represents $\rho_s(T) = 2T/\pi$. (b-d) Scaling behavior of the spin stiffness across a range of temperatures for different magnetic fields. The corresponding magnetic field $h$ and critical temperature $T_c$ are: (b) $h=0.4$, $T_c = 0.149(2)$, (c) $h=1.14$, $T_c = 0.0177(2)$, and (d) $h=1.16$, $T_c = 0$.
        }
    \label{sm_fm_stf}
    \end{figure}

\subsection{C. Details of the Quantum Monte Carlo Methods}

We utilize the sign-free stochastic series expansion (SSE) quantum Monte Carlo (QMC) method with the directed loop update algorithm~\cite{sandvik1991, Sandvik_1992, sandvik2002} to investigate the phase diagram of the model with $K=-2J$ under a magnetic field applied along the $z$-direction. 
In this section, we provide details of the SSE method as applied to this model.

As discussed in the main text, after the transformation, the Hamiltonian takes the form:
$H = K/2 \sum_{\langle ij \rangle}\mathbf{S}_i \cdot \mathbf{S}_j - \sum_i h_i S_i^z,$
where the total number of bonds $\langle ij \rangle$ is denoted by $N_b$, and the total number of sites $i$ is $N_s$. These quantities satisfy the relationship:
$N_b = \frac{N_s \cdot z}{2},$
where the coordination number $z$ is $3$ for the honeycomb lattice.

Therefore, the Hamiltonian can be expressed as:
\begin{equation}
H = \frac{K}{2} \sum_{b=1}^{N_b} \mathbf{S}_{i(b)} \cdot \mathbf{S}_{j(b)} - 
\frac{h}{z} \sum_{b=1}^{N_b} \big[(-1)^{\alpha(b)} S_{i(b)}^z + (-1)^{\beta(b)} S_{j(b)}^z\big],
\end{equation}
where $\alpha(b)$ and $\beta(b)$ correspond to the magnetic field directions of the two sites connected by the bond $b$.

By combining the terms in the Hamiltonian, it can be rewritten as:
\begin{equation}
H = -\sum_{b=1}^{N_b} \big(H_{1,b} - H_{2,b}\big) + N_b \cdot C,
\end{equation}
where $H_{1,b}$ and $H_{2,b}$ represent the contributions from the diagonal and off-diagonal terms, respectively. For the case of $K > 0$, these terms are explicitly given as:
\begin{eqnarray}
H_{1, b}&=&\frac{K}{2}\left(\frac{1}{4}-S_{i(b)}^{z} S_{j(b)}^{z} \right) +\frac{h}{z}\big [(-1)^{\alpha(b)} S_{i(b)}^z+(-1)^{\beta(b)} S_{j(b)}^z +1 \big ]+\epsilon, \\
H_{2, b}&=&\frac{K}{2}\left(\frac{1}{2}S_{i(b)}^{+} S_{j(b)}^{-}+\frac{1}{2}S_{i(b)}^{-} S_{j(b)}^{+}\right), ~~C=\frac{K}{8}+\frac{h}{z}+\epsilon
\end{eqnarray}

Here, $H_{1,b}$ is referred to as the diagonal operator, and $H_{2,b}$ is the off-diagonal operator. The constant $C$ is introduced to ensure that the weight is a positive constant, thereby making the method sign-problem-free.

For the case of $K < 0$, the term $\frac{1}{4}$ in $H_{1,b}$ should be replaced by $-\frac{1}{4}$, and the constant $C$ should be adjusted accordingly.

The partition function can be further expanded as
\begin{equation}\label{eqz}
Z=\sum_{\alpha} \sum_{n=0}^{\infty} \frac{\beta^{n}}{n !}\left\langle\alpha\left|(-H)^{n}\right| \alpha\right\rangle,
\end{equation}
where $(-H)^n$ can be expressed as:
\begin{equation}
(-H)^{n}=\sum_{\left\{H_{a b}\right\}} \prod_{p=1}^{n} H_{a(p), b(p)}.
\end{equation}

We introduce a large number $M$ as the cut-off length. For cases where $n < M$, we insert $(M-n)$ unit operators $H_{0,0} = 1$, leading to:

\begin{equation}\label{hn}
(-H)^{n}=\sum_{\left\{H_{a b}\right\}} \frac{(M-n) ! n !}{M !} \prod_{p=1}^{M} H_{a(p), b(p)}.
\end{equation}

Substituting Eq.\ref{hn} into Eq.\ref{eqz} yields:
\begin{eqnarray}
Z&=&\sum_{\alpha} \sum_{\left\{H_{a b}\right\}} W \left(\alpha,\left\{H_{a b}\right\} \right)\\
W \left(\alpha,\left\{H_{a b}\right\} \right)&=& \frac{\beta^{n}(M-n) !}{M !}\left\langle\alpha\left|\prod_{i=1}^{M} H_{a(i), b(i)}\right| \alpha\right\rangle.
\end{eqnarray}

This formulation enables efficient computation using SSE methods and can be simulated using the directed loop update algorithm \cite{sandvik2002}.

\end{widetext}

\end{document}